\documentclass[aps,onecolumn,prb,showpacs,multicol,amsmath,amssymb]{revtex4}

\usepackage{color}
\usepackage{graphicx}
\usepackage{dcolumn}
\usepackage{bm}

\newcommand{\be}{\begin{equation}}
\newcommand{\ee}{\end{equation}}

\newcommand{\bea}{\begin{eqnarray}}
\newcommand{\eea}{\end{eqnarray}}
\newcommand{\bd}{\begin{displaymath}}
\newcommand{\ed}{\end{displaymath}}
\newcommand{\ba}{\begin{array}}
\newcommand{\ea}{\end{array}}
\newcommand{\bi}{\begin{itemize}}
\newcommand{\ei}{\end{itemize}}
\newcommand{\bc}{\begin{center}}
\newcommand{\ec}{\end{center}}
\newcommand{\bfl}{\begin{flushleft}}
\newcommand{\efl}{\end{flushleft}}
\newcommand{\bfr}{\begin{flushright}}
\newcommand{\efr}{\end{flushright}}

\def\6{\partial}

\def\={\!\!\!&=&\!\!\!}
\def\+{\!\!\!&&\!\!\!+~}
\def\-{\!\!\!&&\!\!\!-~}

\begin{document}

\title[]{Multiorbital Spin Susceptibility in a Magnetically Ordered State - \\Orbital versus Excitonic Spin Density Wave Scenario}
\author {J. Knolle$^{1}$, I. Eremin$^{2}$, and R. Moessner$^{1}$}
 \affiliation{
 $^{1}$Max Planck Institute for the Physics of Complex Systems, D-01187 Dresden, Germany}
 \affiliation {$^2$Institut f\"ur Theoretische Physik III, Ruhr-Universit\"at Bochum, D-44801 Bochum, Germany}

\begin{abstract}
We present a general theory of multiorbital spin waves in magnetically ordered metallic systems. Motivated by the itinerant magnetism of iron-based superconductors, we compare the magnetic excitations for two different scenarios: when the magnetic order either sets in on the on-site orbital level;  or when it appears as an electron-hole pairing between different bands  of electron and hole character. As an example we treat the two-orbital model for iron-based superconductors. For small magnetic moments the spin excitations look similar in both scenarios. Going to larger interactions and larger magnetic moments, the difference between both scenarios becomes striking. While in the excitonic scenario the spin waves form a closed structure over the entire Brillouin zone and the particle-hole continuum is gapped, the spin excitations in the orbital scenario can be treated as spin waves only in a close vicinity to the ordering momenta. The origin of this is a gapless electronic structure with Dirac cones which is a source of large damping. We analyze our results in connection with recent neutron scattering measurements and show that certain features of the orbital scenario with multiple order parameters can be observed experimentally.
\end{abstract}

\date{\today}


\maketitle

\section{Introduction}
In recent years correlated multiorbital electronic systems have attracted a lot of
attention due to a variety of ordering phenomena such as unconventional charge and magnetic ordering in Na$_x$CoO$_2$,  and recently discovered high temperature superconductivity in iron-based superconductors\cite{kamihara}.
Many of these multiorbital systems (MOS) are metallic and have Fermi surfaces (FS) consisting of multiple pockets, which makes an effective description with single band models in principle impossible. Moreover, due to strong hybridization effects, the conducting bands are usually linear combinations of different orbitals. It is therefore desirable to calculate spin excitations in magnetically ordered MOS. 

In iron-based superconductors the role of the underlying orbitals in the formation of the magnetic order and
superconductivity is a subject of ongoing  research. In particular, initial models based on the pure fermiology
and the fact that the FS consist of hole, $\alpha$, and electron, $\beta$, pockets separated by the antiferrromagnetic (AF) wave vectors {\bf Q}$_{1} = (\pi,0)$ and {\bf Q}$_{2} = (0,\pi)$ predicted a  superconducting instability with a so-called $s^{+-}-$wave symmetry of the superconducting gap\cite{mazin,chubukov,tesanovic,honerkamp,schmalian}. Correspondingly, the  spin density wave order originates from nesting of the $\alpha$ and $\beta$ bands which yield a logarithmic instability of the spin response at the AF wave vector\cite{chubukov,vavilov,eremin,tesanovic,brydon,knolle}. Within this scenario, often called excitonic, the role of the lattice and orbitals is negligible as the main effect is associated with the electron and hole character of the bands involved. Moreover, the RG studies support this indirectly as the bare local interactions renormalize drastically during the RG flow as a result of the interfering logarithms in the particle-hole and particle-particle channels\cite{chubukov,honerkamp}. Note that despite its simple form, the excitonic scenario describes well spin wave excitations below T$_N$, their dispersions and anisotropy of the spin wave velocities\cite{knolle}.

At the same time, it was argued that the orbitals do play an important role, namely, by introducing a significant angular dependence for the intraband and interband interactions resulting in a  change of important details in the superconducting and antiferromagnetic states. For example, it was argued that the originally nodeless $s^{+-}-$ superconducting state  may acquire nodes on the electron pockets\cite{vavilov1,graser,thomale,arita}. For the AF state it was predicted that iron-based superconductors show nodal elementary excitations with a linear Dirac spectrum at low energies as a result of orbital effects\cite{vishwanath,kim,tohyama}.  In addition, due to the fact that the antiferromagnetic order is either $(\pi,0)$ or $(0,\pi)$, the magnetic anisotropy induces orbital anisotropies which in turn affect the electronic properties along $a$ and $b$ directions\cite{tohyama1,daghofer,leni}. They may also introduce novel features such as ferro-orbital or antiferro-orbital order\cite{Thalmeier2008,Lv} associated with intra and inter-orbital electron-hole pairing. Therefore, excitonic band and orbital lattice descriptions are not equivalent even in the limit of interactions being much smaller than the bandwidth. Furthermore, nesting features do not play an essential role  in the formation of the AF order starting from the orbital description.

In this paper we analyze itinerant spin excitations of the AF state with either $(\pi,0)$ or $(0,\pi)$ ordering wave vector for the two orbital model in the excitonic and orbital scenarios.
The aim of the present work is two fold: on the one hand we want to present the general spin wave theory in MOS with itinerant magnetism in a self-consistent random phase approximation (RPA)-type formalism. We compare in detail the excitonic and orbital AF scenario and highlight important differences in the properties of the magnetic excitations in both cases. On the other hand, we analyze the magnetic state of iron-based superconductors. In particular, we show that for small and intermediate values of the magnetic moment, spin waves in the
excitonic scenario have longer life times while the low-energy continuum reveals stronger damping of them in the orbital description.  We further analyze the magnetic excitations of the two orbital model and discuss its connection to recent experiments. In the orbital resolved susceptibilities, we observe major differences between the spin waves coming from different orbitals and discuss its relevance for resonant inelastic X-ray scattering (RIXS).

The paper is organized in the following way: After giving the basic definitions we review the multiorbital formalism for calculating RPA spin susceptibilities in the paramagnetic phase and extend it to the symmetry broken AF state in a self-consistent way. Using as an example the two orbital model\cite{Raghu2008} we provide explicit expressions of the susceptibility tensor and interaction matrix and explain in detail the differences between the excitonic and orbital AF scenarios. Although we treat a concrete two orbital model, the formalism is easily applicable to different MOS. In the {\it Results} section we present a comparison between the orbital and excitonic spin waves, their damping, and evolution from small to large magnetic moments. Finally, we show the orbital resolved susceptibilities and discuss the possibility of observing them in RIXS experiments and make connection to recent INS experiments. We close the paper with a {\it Summary} section.

Our results for iron-based superconductors are complementary to previous studies. For example, Ref.\onlinecite{Tohyama_spinwaves} discusses spin waves in iron-based superconductors using the orbital picture within a five orbital model. The collective spin excitations in SDW phases using simplified band models have been calculated in Refs.\onlinecite{brydon,knolle}. Here, we analyze both orbital and band descriptions within the same two orbital model and critically compare both scenarios. We also provide an extensive discussion of the method which can easily be implemented for other multiorbital models.

\section{Theory}
We begin by discussing general aspect of the self-consistent
RPA-type theory of spin excitations in the AF ordered phase.
For a single band model the formalism was presented in Refs.\onlinecite{Schrieffer1989,frenkel}. Here, we demonstrate the theory in detail for a two orbital model in the excitonic and orbital AF scenarios using the model Hamiltonain proposed by Raghu et al.\cite{Raghu2008} for iron based superconductors. The generalization to different MOS and models with more orbitals is straightforward.

\subsection{Basic Definitions}
We define the spin operator as
\begin{eqnarray}
\label{Spin}
{\bf S}_s({\bf q})  & = &  \sum_{\mathbf{k}, \sigma, \sigma'} \psi_{s\sigma}^{\dagger}({\bf k+q}) \frac{{\vec \sigma}_{\sigma \sigma'}}{2} \psi_{s\sigma'}({\bf k}) ,
\end{eqnarray}
where $\psi^{\dagger}_{s\sigma}$ is the creation operator of the orbital $s$ with spin $\sigma$ in the MOS. The transverse spin susceptibility can be calculated from the spin-spin correlation function in Matsubara frequency space:
\begin{eqnarray}
\label{SpinCorr}
\chi_{st}^{\pm}({\bf q}, i\Omega)  & = &  \int_{0}^{\beta} d\tau e^{i\Omega \tau} \left\langle T_{\tau} S_{s}^{+}({\bf q}, \tau) S_{t}^{-}({\bf -q}, 0) \right\rangle.
\end{eqnarray}
In MOS it is convenient to introduce a dynamical spin susceptibility tensor in a more general way using four orbital indices
\begin{eqnarray}
\label{SpinCorr2}
\chi_{spqt}^{\pm}({\bf q}, i\Omega)  & = &  \int_{0}^{\beta} d\tau e^{i\Omega \tau}  \sum_{{\bf k}, {\bf k'}, \gamma, \delta, \gamma', \delta'}\left\langle T_{\tau}
\psi^{\dagger}_{p\gamma}({\bf k}, \tau) \psi_{q\delta}({\bf k+q}, \tau) \psi^{\dagger}_{t\gamma'}({\bf k'}, 0) \psi_{s\delta'}({\bf k'-q}, 0) \right\rangle \sigma_{\gamma \delta}^{+}\sigma_{\gamma' \delta'}^{-}.
\end{eqnarray}
With the definition of the Green's function (GF)
\begin{eqnarray}
\label{GF}
G_{st}^{\sigma}({\bf k}, i\omega_n)  & = &  -\int_{0}^{\beta} d\tau e^{i\omega \tau}  \left\langle T_{\tau} \psi_{s\sigma}({\bf k}, \tau) \psi_{t\sigma}^{\dagger}({\bf k}, 0) \right\rangle
\end{eqnarray}
the susceptibility matrix can be written as
\begin{eqnarray}
\label{SpinCorr3}
\chi_{spqt}^{\pm}({\bf q}, i\Omega)  & = &  -\frac{1}{N\beta} \sum_{{\bf k}, i\omega_n} G_{sp}^{\uparrow}({\bf k}, i\omega_n) G_{qt}^{\downarrow} ({\bf k+q}, i\omega_n+i\Omega).
\end{eqnarray}
The susceptibility in the paramagnetic phase within the RPA can be written in a matrix form $\hat\chi$, see  for example Ref.\onlinecite{graser}
\begin{equation}
\label{RPA}
\chi_{bsta}^{RPA} =\chi_{bsta}^0+\chi_{b'sta'}^{RPA}  U_{c'b'a'd'} \chi_{bc'd'a}^0 \quad,
\end{equation}
and the solution of the RPA in matrix form is straightforward.
\begin{equation}
\left[\hat\chi\right]^{RPA} =\hat\chi^0 (1-\hat U \hat\chi^0)^{-1}
\end{equation}
The physical susceptibility which is measured in experiment is then obtained
\begin{equation}
\chi_{phys}^{RPA}({\bf q},\omega)=\frac{1}{2} \sum_{st} \chi^{RPA}_{sstt}({\bf q},\omega).
\end{equation}

Generally, for MOS, the kinetic or tight-binding part $H_0$ of the Hamiltonian $H=H_0+H_{int}$, including the intra- and inter-orbital hoppings, can be diagonalized via a unitary transformations which gives the eigenergies $E_{\nu}({\bf k})$ of the system. The operators in the orbital and the band basis are connected via the coefficients of the unitary transformations $a_{\nu}^{s}({\bf k,\sigma})$
\begin{equation}
\label{Diag}
\psi_{s\sigma}({\bf k}) = \sum_{\nu} a_{\nu}^{s}({\bf k,\sigma}) \gamma_{\nu \sigma} ({\bf k})
\end{equation}
Without interactions the GFs are easily obtained as
\begin{eqnarray}
\label{GFcalc}
G_{st}^{\sigma}({\bf k}, i\omega_n)  & = &  \sum_{\nu}\frac{a_{\nu}^{s}({\bf k, \sigma}) a_{\nu}^{t *}({\bf k, \sigma}) }{i\omega -E_{\nu} ({\bf k})}
\end{eqnarray}
and the transverse spin susceptibility becomes:
\begin{eqnarray}
\label{SpinCorrcalc}
\chi_{spqt}^{\pm}({\bf q}, i\Omega)  & = &  -\frac{1}{N\beta} \sum_{{\bf k}, i\omega} \sum_{\nu, \mu} \frac{a_{\mu}^{s}({\bf k, \uparrow}) a_{\mu}^{p *}({\bf k, \uparrow}) a_{\nu}^{q}({\bf k+q, \downarrow}) a_{\nu}^{t *}({\bf k+q, \downarrow}) }{\left[ i\omega-E_{\mu}({\bf k})\right] \left[ i\omega+i\Omega -E_{\nu}({\bf k+q})\right] }.
\end{eqnarray}
The Matsubara sum can be performed via a standard contour integration and by analytic continuation $i\Omega \rightarrow \Omega+i\Gamma$. Observe that for large Hamiltonian matrices the coefficients $a_{\nu}$ have to be calculated numerically by diagonalizing the Hamiltonian matrix written in {\bf k}-space. In the following we move to the AF state.

In a mean-field treatment of the SDW phase, the interaction part of the hamiltonian $H_{int}$ is decoupled with respect to a groundstate with a finite magnetization at wave vector ${\bf Q}$, e.g. $(\pi,0)$ or $(0,\pi)$ for iron-based superconductors. The order parameters read
\begin{eqnarray}
 \sigma \Delta_{\nu\nu'} & = & \sum_{{\bf k}} \left\langle \psi_{\nu \sigma}^{\dagger} ({\bf k}) \psi_{\nu' \sigma} ({\bf k+Q})\right\rangle
\end{eqnarray}
where $\sigma=\pm 1$ refers to up or down spins.
In the magnetic state, SU(2) spin symmetry is broken which yields
$\chi^{\pm}  \neq  \frac{1}{2}\chi^{zz}$. In addition, the AF state also breaks translational symmetry and the mean-field Hamiltonian in the reduced BZ takes the form
\begin{eqnarray}
H_{SDW} & = & \sum_{{\bf k}}^{'} \hat\psi_{\sigma}^{\dagger} ({\bf k})
\begin{pmatrix}
 H_0 ({\bf k}) &  \sigma  M \\
 \sigma  M &  H_0 ({\bf k+Q})
\end{pmatrix}
\hat \psi_{\sigma} ({\bf k})
\end{eqnarray}
where  we introduce the new spinor
\begin{eqnarray}
\label{SpinAFM}
 \hat\psi_{\sigma}({\bf k}) & = &
\begin{pmatrix}
\psi_{\sigma} ({\bf k}) \\
\psi_{ \sigma} ({\bf k+Q})
\end{pmatrix}.
\end{eqnarray}
Note for MOS each component of the spinor may acquire a spinor structure in a orbital space as well.
Then the GF acquires off-diagonal elements in the momentum space and becomes a matrix. In short-hand notation, we write only the orbital index with a tilde to indicate that the momentum is shifted by the AFM wave vector. We thus obtain
\begin{eqnarray}
\hat G_{sp} & = & \left\langle
\begin{pmatrix}
s \\
\tilde s
\end{pmatrix}
\begin{pmatrix}
p^{\dagger} & \tilde p^{\dagger}
\end{pmatrix}
\right\rangle
=
\begin{pmatrix}
G_{sp} & G_{s\tilde p} \\
G_{\tilde s p} & G_{\tilde s \tilde p}
\end{pmatrix}
\end{eqnarray}
Analogously, the susceptibility matrix becomes a matrix in ${\bf q}$-space\cite{Schrieffer1989,brydon}:
\begin{eqnarray}
 \hat \chi_{spqt} ^{\pm} & = &
\begin{pmatrix}
\chi_{spqt} ^{\pm}({\bf q},{\bf q}) & \chi_{spqt} ^{\pm}({\bf q},{\bf q+Q})  \\
\chi_{spqt} ^{\pm}({\bf q+Q},{\bf q}) & \chi_{spqt} ^{\pm}({\bf q+Q},{\bf q+Q})
\end{pmatrix}\\
& = &
\label{SuscLoopAFM}
\begin{pmatrix}
\left[ G_{sp}G_{qt}+G_{s\tilde p}G_{\tilde q t} + G_{\tilde s \tilde p}G_{\tilde q \tilde t}+G_{\tilde s p}G_{q \tilde t} \right] & \left[ G_{sp}G_{q\tilde t}+G_{s\tilde p}G_{\tilde q \tilde t} + G_{\tilde s \tilde p}G_{\tilde q  t}+G_{\tilde s p}G_{q t} \right] \\
\left[ G_{sp}G_{q\tilde t}+G_{s\tilde p}G_{\tilde q \tilde t} + G_{\tilde s \tilde p}G_{\tilde q  t}+G_{\tilde s p}G_{q t} \right]  &\left[  G_{ s p}G_{\tilde q\tilde t}+G_{ s\tilde p}G_{ q \tilde t} + G_{\tilde s  p}G_{\tilde q  t}+G_{\tilde s\tilde p}G_{ q t} \right]
\end{pmatrix}
\label{SuscMatrixAF}
\end{eqnarray}
where the off-diagonal components are called Umklapp susceptibilities as they connect spins with momentum shifted by the AFM wave vector, which is a reciprocal wave vector of the magnetic BZ.

The form of the RPA in the SDW state remains the same, see Eq.(\ref{RPA}), with the only extension that the
interaction matrix also doubles in size with $\hat U$ matrices on the diagonal and zeros on the off-diagonal. To obtain the physical susceptibility we perform an RPA with Eq.(\ref{SuscMatrixAF}) and trace over the orbital indices of the first entry $\chi_{spqt} ^{RPA}({\bf q},{\bf q})$ of the susceptibility matrix. The imaginary part is directly related to the cross section of INS measurements. We also recall that self-consistency enters via taking the same set of interaction parameters in the RPA and the Hartree-Fock mean-field equations.

\subsection{Two orbital model: excitonic versus orbital SDW scenario}
\begin{figure}[t]
\centering
\includegraphics[width=1.0\textwidth]{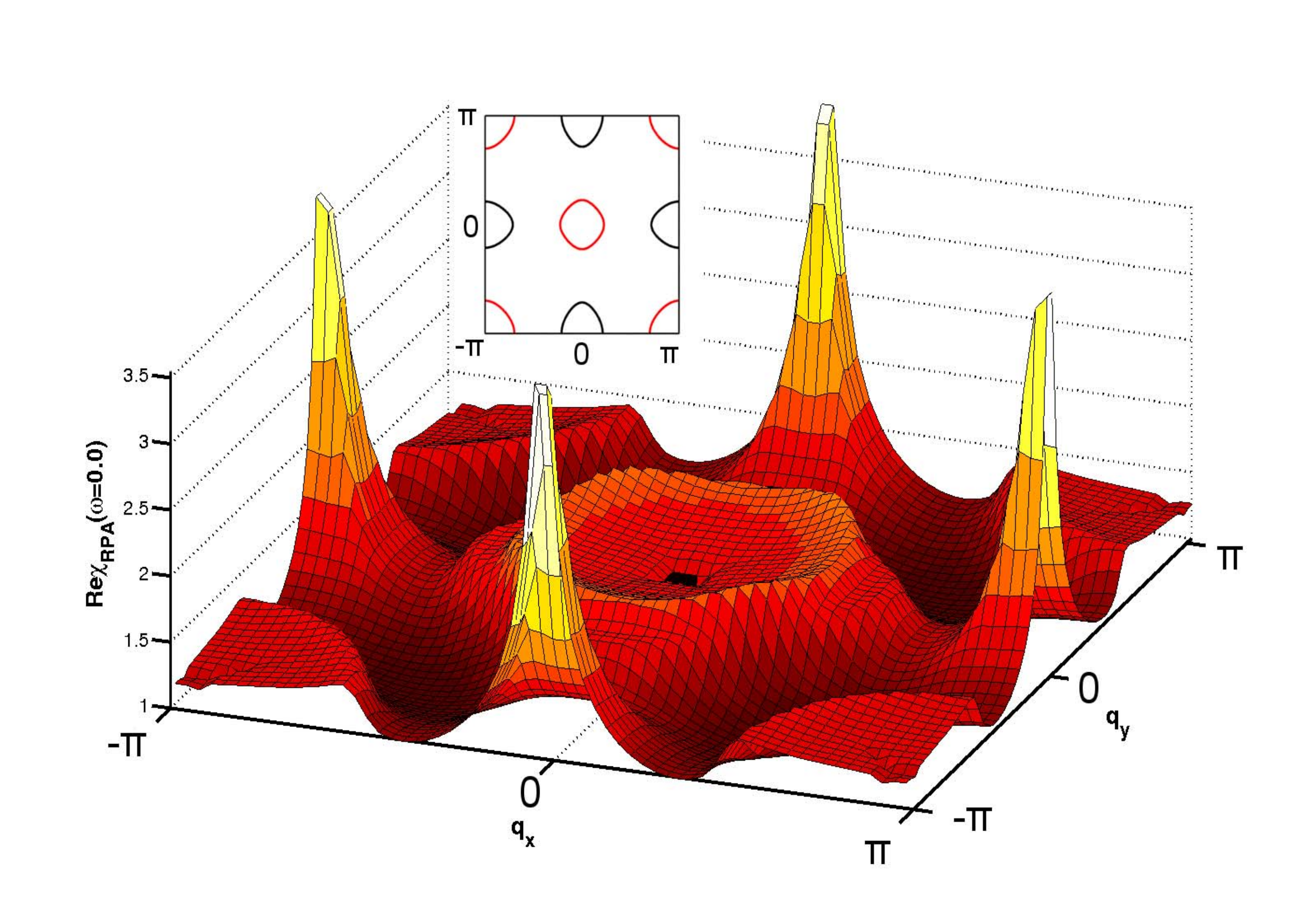}
\caption{Calculated real part of the physical susceptibility for the two-orbital model. The inset shows the corresponding Fermi surface topology. Here, we use  $t_1=-1,t_2 = 1.3,
t_3=t_4=-0.85$, in units of $\vert t_1 \vert = t$.}
\label{fig1}
\end{figure}
Let us illustrate the derivation, presented above, for the two orbital model\cite{Raghu2008} with $d_{xz}$ and $d_{yz}$ orbitals  proposed for the iron-based superconductors. The kinetic part of the Hamiltonian is written in the two component spinor basis in orbital space with the help of the Pauli matrices $\tau_i$
\begin{eqnarray}
\label{Hraghu}
H_0 & = &  \sum_{{\bf k},\sigma }  \psi_{\sigma}^{\dagger}({\bf k}) \left[ \left( \epsilon_{+} ({\bf k})-\mu \right) \tau_0 + \epsilon_{-} ({\bf k}) \tau_3 +\epsilon_{xy} ({\bf k}) \tau_1\right]  \psi_{\sigma}({\bf k})
\end{eqnarray}
where
\begin{eqnarray}
\label{Spinor}
 \psi_{\sigma}({\bf k}) & = &
\begin{pmatrix}
d_{xz \sigma} ({\bf k}) \\
d_{yz \sigma} ({\bf k})
\end{pmatrix}
\end{eqnarray}
Here, $\varepsilon_\pm(k)  =  \frac{\varepsilon_x(k)\pm\varepsilon_y(k)}{2}$, $\varepsilon_x(k)  =  -2t_1\cos k_x-2t_2\cos k_y-4t_3\cos k_x\cos k_y$, $\varepsilon_y(k) =  -2t_2\cos k_x-2t_1\cos k_y-4t_3\cos k_x\cos k_y$, and
$\varepsilon_{xy}(k)  =  -4t_4\sin k_x\sin k_y$. The resulting Fermi surface is shown in the inset of Fig.\ref{fig1}.

The interacting part of the Hamiltonian consists of different on-site electron-electron interactions:
\begin{eqnarray}
\label{HraghuInt}
H^{orb}_{int} & = &  U \sum_{i } \sum_{\nu} n_{i\nu\uparrow}n_{i\nu\downarrow} +V \sum_{\nu \neq \mu,\sigma,\sigma'}
n_{i\nu\sigma}n_{i\mu\sigma'} -J \sum_{\nu \neq \mu} {\bf S}_{i\nu}\cdot {\bf S}_{i\mu} +J^{\prime}\sum_{\nu\neq\mu} d_{i\nu\uparrow}^{\dagger} d_{i\nu\downarrow}^{\dagger}d_{i\mu\downarrow}d_{i\mu\uparrow}.
\end{eqnarray}
where $U$ and $V$ refer to the intra- and inter-orbital Coulomb repulsion, $J$ and $J^{\prime}=J/2$ denote the Hund and the pair hopping terms.
The susceptibility matrix, with labels $1$ for $xz$ and $2$ for $yz$,  reads
\begin{equation}
\label{SuscMatrix}
\hat\chi=
\begin{bmatrix}
\chi_{1111} & \chi_{2111} & \chi_{1112} & \chi_{2112}  \\
\chi_{1211} & \chi_{2211} & \chi_{1212} & \chi_{2212}  \\
\chi_{1121} & \chi_{2121} & \chi_{1122} & \chi_{2122}  \\
\chi_{1221} & \chi_{2221} & \chi_{1222} & \chi_{2222}  \\
\end{bmatrix}
\end{equation}
and the interaction matrix for the transverse component of the spin susceptibility is given by:
\begin{equation}
\label{IntMatrix}
\hat U=
\begin{bmatrix}
U & 0 & 0 & J  \\
0 & J & U-2J & 0  \\
0 & U-2J & J & 0  \\
J & 0 & 0 &U  \\
\end{bmatrix}.
\end{equation}
Following the procedure described above it is straightforward to compute the physical susceptibility in the paramagnetic state. In particular, Fig.\ref{fig1} shows the real part of the physical susceptibility for the two orbital model. Observe that in the paramagnetic state the response is governed by the antiferromagnetic fluctuations centered at $(\pi,0)$ and $(0,\pi)$.

In the next step we introduce self-consistent mean-field averages associated with the AF state. As described above, we consider two scenarios.

In the first case, which we denote as {\it orbital},  one introduces the magnetic state in the orbital basis using the local interactions. The mean-field phase diagram of the two orbital model was studied extensively \cite{vishwanath,Thalmeier2008} and the main conclusion is that AF ordering involves only very weakly interorbital components. Thus we assume in the following that the mean-field AF order involves only states within the same orbital, separated by ${\bf Q}_{AF}$ ($\Delta_{\nu \nu'}=\delta_{\nu,\nu'} \Delta_{\nu}$), and decouple the on-site local interaction, Eq.(\ref{HraghuInt}), with respect to it. In addition, we set ${\bf Q}_{AF}=(0,\pi)$ and do not consider the problem of its selection out of the manifold of degenerate magnetic states as it was considered by us previously\cite{eremin}. We obtain  two coupled gap  equations
\begin{eqnarray}
\label{SelfConsGapOrb}
M_{x} & = & - \left[ U \Delta_{xz} + J \Delta_{yz} \right] \\
M_{y} & = & - \left[ U \Delta_{yz} + J \Delta_{xz} \right] \\
\end{eqnarray}
and the Hamiltonian becomes
\begin{eqnarray}
\hat h^{Orbital} & = &
\begin{pmatrix}
\varepsilon_x & \varepsilon_{xy} & \sigma M_x & 0 \\
\varepsilon_{xy} & \varepsilon_y & 0 & \sigma M_y \\
\sigma M_x & 0 & \tilde \varepsilon_x & \tilde \varepsilon_{xy} \\
0 & \sigma M_y & \tilde \varepsilon_{xy} & \tilde \varepsilon_y
\end{pmatrix}
\end{eqnarray}
where the tilde again means that the momentum is shifted by ${\bf Q}_{AF}$
The matrix can be inverted and with the new eigenvectors one can solve the coupled gap equations and carry out the spin wave calculations as outlined above.

The second, {\it excitonic}, scenario is built on the belief that  in the iron-based superconductors,  the antiferromagnetic instability, at least partially, is connected to  nesting of electron  and hole bands\cite{mazin,eremin}. In this scenario, the $\log$ instability of the particle-hole response function is similar to that in the particle-particle Cooper-channel. In this regard it seems natural to assume that the magnetic instability arises not in the orbital representation but on the band level for electrons lying close to the Fermi level. This situation is not unique to ferropnictides and is believed to arise also in Cr and is sometimes dubbed  excitonic antiferromagnetism. Within this scenario the tight-binding hamiltonian,Eq.(\ref{Hraghu}), is first diagonalized to obtain the band structure. One further assumes, that the bands are the relevant degrees of freedom and ignores the role of the underlying orbital structure. The interactions are then also re-written in the band representation. Moreover, within the RG flow they are strongly renormalized, to such an extent that their values at the Fermi energy are not the same as the initial ones determined by the local on-site interactions\cite{chubukov,maiti}. This allows us to approximate the band interactions as Fermi-liquid density-density type interactions, which at the lowest order are angular independent\cite{chubukov}. Following this we consider the density-density interactions which give rise to SDW ordering at $(\pi,0)$ or $(0,\pi)$ \cite{footnote1}:
\begin{eqnarray}
\label{BandInt}
H_{int}^{band} & = & \sum_{\begin{matrix} {\bf k, k', q} \\ \sigma, \sigma' \end{matrix}} u_1 \left[ \gamma_{1\sigma}^{\dagger} ({\bf k+q}) \gamma_{2\sigma'}^{\dagger} ({\bf k'-q}) \gamma_{2\sigma'} ({\bf k'}) \gamma_{1\sigma} ({\bf k}) \right] + \frac{u_3}{2} \left[ \gamma_{1\sigma}^{\dagger} ({\bf k+q}) \gamma_{1\sigma'}^{\dagger} ({\bf k'-q}) \gamma_{2\sigma'} ({\bf k'}) \gamma_{2\sigma} ({\bf k}) + h.c.\right]
\end{eqnarray}
The $\gamma$-operators, see Eq.(\ref{Diag}), are written in the band representation which diagonalizes the tight-binding part of the Hamiltonian with Eigenenergies $E_{\nu}({\bf k})$.

In the excitonic scenario the AF order involves quasiparticle states from the electron and hole bands connected by the AF momentum ${\bf Q_1}=(0,\pi)$. By decoupling the band interaction with respect to $\Delta_{12}$ one finds a single gap equation
\begin{eqnarray}
\label{SelfConsGapExc}
M & = & -\frac{u_1+u_3}{2} \Delta_{12} \\
\end{eqnarray}
and the Hamiltonian matrix becomes
\begin{eqnarray}
\hat h^{Band} & = &
\begin{pmatrix}
E_1 & 0 & 0 & \sigma M \\
0 & E_2 & \sigma M & 0 \\
0 & \sigma M & \tilde E_1 & 0 \\
\sigma M & 0 & 0 & \tilde E_2
\end{pmatrix}.
\end{eqnarray}
\begin{figure}[]
\centering
\includegraphics[width=1.0\textwidth]{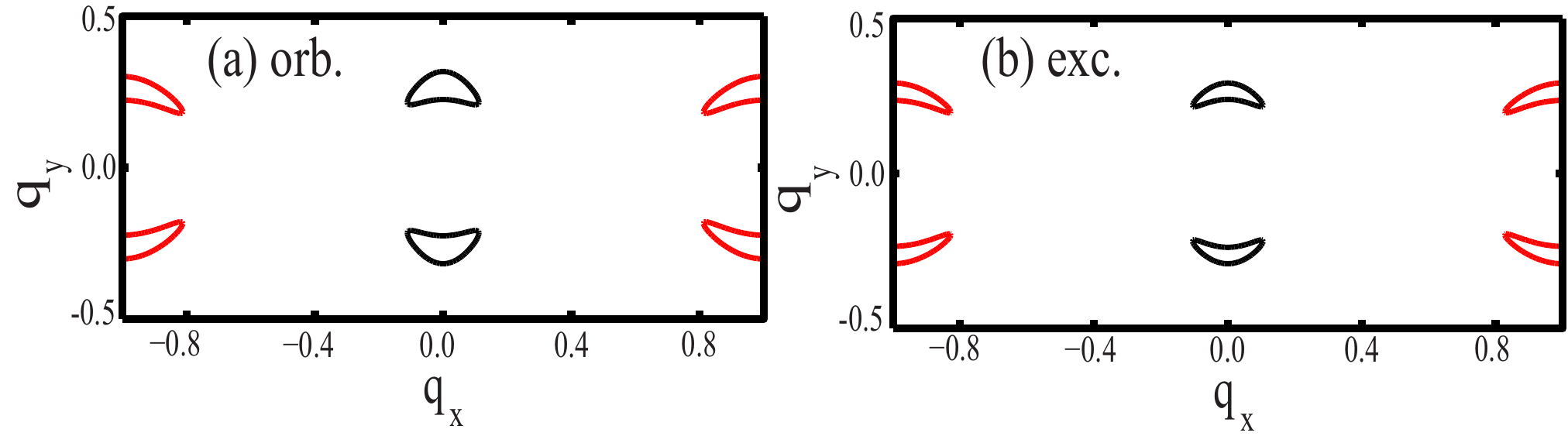}
\caption{Fermi surface of the AF state with a magnetic moment of $m=0.23\mu_B$ in the orbital (a) and excitonic (b) scenarios. Despite the fact that the FS topology is similar in both cases its origin is different. While in the excitonic scenario the remnant FS pockets arise due to imperfect nesting between electron and hole bands and a small size of the SDW gap, the electronic structure and the Fermi surface pockets in the orbital scenario are consequences of the nodal SDW\cite{vishwanath} and the pockets do not vanish for increasing the size of the SDW gaps but simply shift their position in the reduced BZ.}
\label{fig2}
\end{figure}
Within the excitonic scenario we obtain a constant gap splitting of the excitation spectra all over the first BZ.
By contrast, in the orbital picture the overall gap in the excitation spectra appears to be nodal as a result of the underlying symmetry of the orbital structure\cite{vishwanath}. In addition, the structure  of the AF state is richer as it involves at least two intra-orbital SDW orders. Despite these differences between the orbital and excitonic scenario, it is interesting to note that the resulting FSs for small sizes of the magnetic moments appear to be quite similar as shown in Fig.\ref{fig2}. Nevertheless, the origin of the pockets is different.  In the excitonic scenario the remnant FS pockets arise due to imperfect nesting between electron and hole bands and a small size of the SDW gap. On the contrary, the Fermi surface pockets in the orbital scenario are a  consequence of the nodal SDW\cite{vishwanath} and they do not vanish for increasing size of the SDW gaps but simply shift their position in the reduced BZ.
The difference between both scenarios becomes apparent with increasing size of the magnetic moments as, we show below.
\begin{figure}[]
\centering
\includegraphics[width=0.5\textwidth]{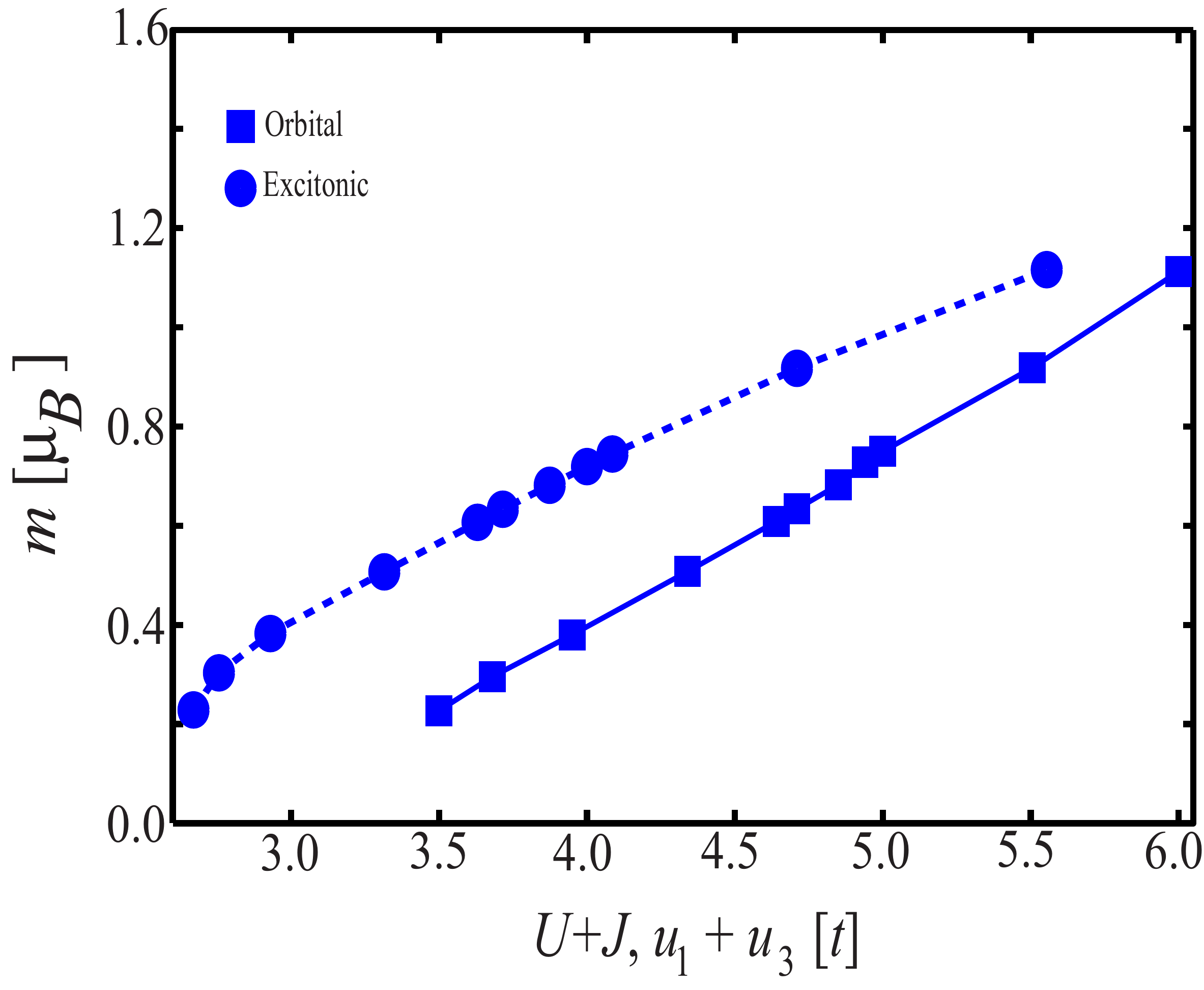}
\caption{Calculated sizes of the magnetic moments in the excitonic and orbital scenarios as a function of the interaction strength. The solid and the dashed lines are guides to the eye.}
\label{figmm}
\end{figure}

In Fig.\ref{figmm} we show the evolution of the magnetic moment within the orbital and excitonic scenario with increasing interaction strength ($U+J$ and $u_1 +u_3$ respectively). Observe that the excitonic scenario requires a smaller value of the interaction to obtain sizeable SDW gaps and the corresponding magnetic moments. This is a result of neglecting orbital matrix elements which overall tend to weaken the effect of interactions in the SDW gap equation. Indeed within an excitonic scenario, the SDW gap  and the magnetic moment appear for smaller magnitude of the  interactions as compared to the orbital case. The origin is the logarithm of the particle-hole channel in the excitonic scenario. Although, also present in the orbital scenario, its effect is weakened by the underlying orbital character of the wave functions. Thus within the orbital scenario the SDW gaps increase much slower with increase of the moment. In addition for $\mu \geq 0.38\mu_B$ the Fermi surface is completely gapped within the excitonic scenario, while FS pockets are always present in the orbital scenario. The latter leads to finite damping of the spin waves;  well-defined dispersive spin waves are thus absent from much of the BZ in the orbital scenario as will be shown in the following.

\section{Results}
In the following we analyze spin excitations within the excitonic and orbital scenarios. The interaction parameters -- $U$ and $J$ for the orbital scenario, and $u_1+u_3$ for the excitonic scenario -- determine the magnitude of the magnetic moment via the self-consistent gap equations, Eqs.(\ref{SelfConsGapExc} and \ref{SelfConsGapOrb}). We always choose them  in such a way that the resulting moment is equal within both scenarios which allows a direct comparison of the spin waves. We concentrate on the undoped case  that corresponds to a total number of particles $n=2$ per unit cell. Observe also that in the orbital scenario occupation and magnetization of the two orbitals are unequal because the AFM order with {\bf Q}$_1$=$(0,\pi)$ ordering momentum breaks the symmetry between the two orbitals.

\begin{figure}[]
\centering
\includegraphics[width=1.0\textwidth]{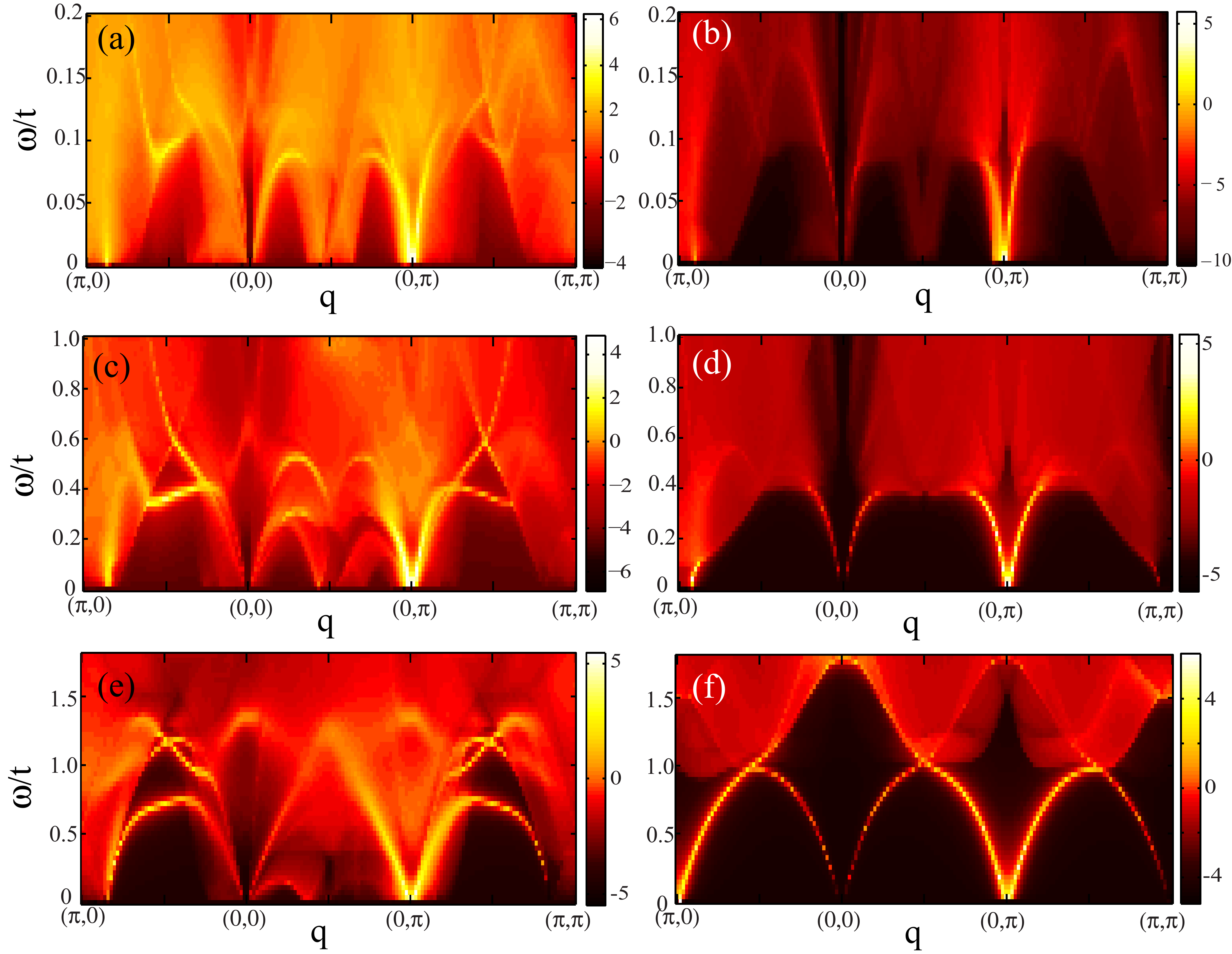}
\caption{Calculation of the imaginary part of the transverse spin susceptibility in the orbital (left panel) and excitonic (right panel) scenarios for the magnetic moment of $m=0.23 \mu_B$(a)-(b), $m=0.38 \mu_B$(c)-(d) and $m=0.75 \mu_B$(e)-(f). The intensity in states/eV is shown on a log scale. For concreteness, we set $J=0.1U$ and use a quasiparticle damping of $\Gamma=$0.005t.}
\label{fig3}
\end{figure}

In Fig.\ref{fig3} we show the evolution of the transverse spin excitations in the orbital (left) and excitonic (right) scenarios along the route $(\pi,0) \rightarrow (0,0) \rightarrow (0,\pi) \rightarrow (\pi,\pi)$ of the first BZ and for various values of the magnetic moment. The intensity is shown on a log intensity scale. In panels (a)-(b) the magnetic moment is small, $m=0.23 \mu_B$. In the excitonic case we use  $\mu=1.54t$, $(u_1+u_3)/2=1.335 t$ which gives a gap of $\Delta_{12}=0.1525 t$, while in the orbital case we employ $U=3.5 t$  and the gaps are $\Delta_{xz}=0.2086 t$ and $\Delta_{yz}=0.2283t$. As mentioned above, larger magnitudes of the gaps  are required in the orbital case.

The peculiar feature of the spin excitations in the AF state within both scenarios are the well-defined spin wave excitations around the ordering wave vector {\bf Q}$_{1} = (0,\pi)$ and their visible dispersion up to energies of about $0.1 t$ despite the finite damping originating from remnant FS pockets. At higher energies the spin waves are overdamped by the particle-hole continuum and evolve further as Stoner-like excitations (paramagnons). Another common element of both scenarios for this small value of the magnetic moment is that the particle-hole continuum dominates the low-energy excitations in the entire BZ except around the ordering wave vector itself. Although the details of the continuum slightly differ in both scenarios, the gross features are the same.
In particular, the low-energy excitations around $(0,\pi/2)$, $(\pi,\pi)$, and $(\pi,0)$ are due to particle-hole scattering between remnant parts of the Fermi surface, see Fig.\ref{fig2}. The higher energy features look slightly different due to the richer structure of the particle-hole continuum in the orbital scenario, but their intensity is overall low.

\begin{figure}[h]
\centering
\includegraphics[width=1.0\textwidth]{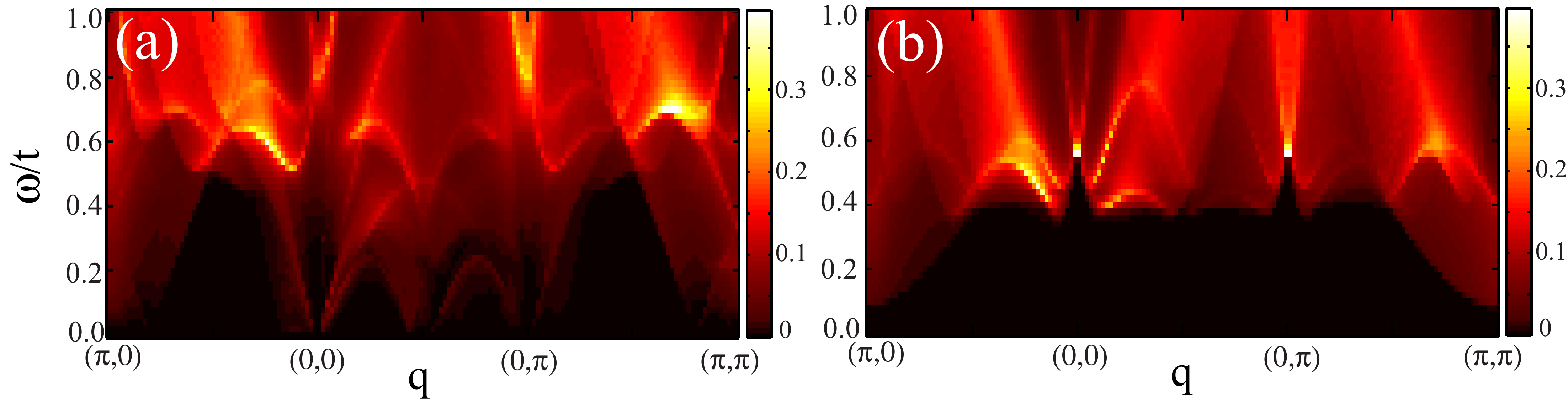}
\caption{Imaginary part of the bare transverse physical susceptibility, Im$\chi^{0}$, for $m=0.38 \mu_B$ for the orbital (a) and excitonic (b) scenario with $m=0.38 \mu_B$.}
\label{fig4}
\end{figure}

The difference between both scenarios becomes more apparent if one raises the moment towards $m=0.38 \mu_B$, see Fig.\ref{fig3}(c)-(d). Here, we use $\mu=1.53t$ and $(u_1+u_3)/2=1.465 t$ with the corresponding SDW gap of $\Delta_{12}=0.2803 t$, while in the other case one needs $U=3.95 t$  and $\Delta_{xz}=0.4008 t$ and $\Delta_{yz}=0.4250 t$. Observe that in this case the continuum of the spin excitations is completely gapped  in the excitonic scenario, as shown also in Fig.\ref{fig4}(b). This is a result of the fully gapped Fermi surface in this case. As a consequence, the spin wave branch is well-defined over almost the entire range of the calculated momenta.

It is interesting to note that within the two orbital model and excitonic scenario of the SDW there is an $O(6)$ degeneracy of the magnetic ground state\cite{eremin} which manifests itself in additional zero energy modes at the non-ordered momenta of $(\pi,0)$ and $(\pi,\pi)$ as shown in Fig.\ref{fig3}(d). These modes, however, are accidental in the sense that they are not stable  with regard to perturbations such as the finite Landau damping introduced by the remnant Fermi surfaces. This is why they are not seen for $m=0.28 \mu_B$. However, as soon as the FSs are completely gapped they become visible.
In the orbital scenario, however, the FSs will never be gapped completely for symmetry reasons and the particle-hole continuum remains gapless as shown in Fig.\ref{fig4}(a). Thus, the spin waves remain well-defined only around the ordering momentum. Away from it they are overdamped because the Dirac cones in the electronic excitation spectra lead to remnant FS pockets.

The situation does not change significantly upon further increase of the interaction. In particular, in Fig.\ref{fig3}(e)-(f) we show the results for the transverse part of the spin susceptibility for the magnetic moment of $m=0.75 \mu_B$. This magnetic moment arises for $\mu=1.53t$ and $(u_1+u_3)/2=2.043 t$ with $\Delta_{12}=0.7607 t$ in the excitonic scenario and for $U=5.0 t$, $\Delta_{xz}=0.9846 t$ and $\Delta_{yz}=1.0772 t$ in the orbital ones. The essential features of the spin waves in the orbital and exciton scenario remain the same. While in the orbital picture the spin waves are overdamped away from the ordering momentum, the spin waves in the excitonic scenario form a closed structure over the entire BZ. Observe also that within both scenarios the spin waves show an anisotropy along the antiferromagnetic ($v_y$) and ferromagnetic ($v_x$) directions. Namely we we find that within both scenarios
the anisotropy is relatively weak, $v_y/v_x \approx 1.04$ in the excitonic and $v_y/v_x \approx 1.06$ in the orbital scenario. This agrees with the result of the $J_1-J_2$-model\cite{Thalmeier2010}.

\begin{figure}[t]
\centering
\includegraphics[width=1.0\textwidth]{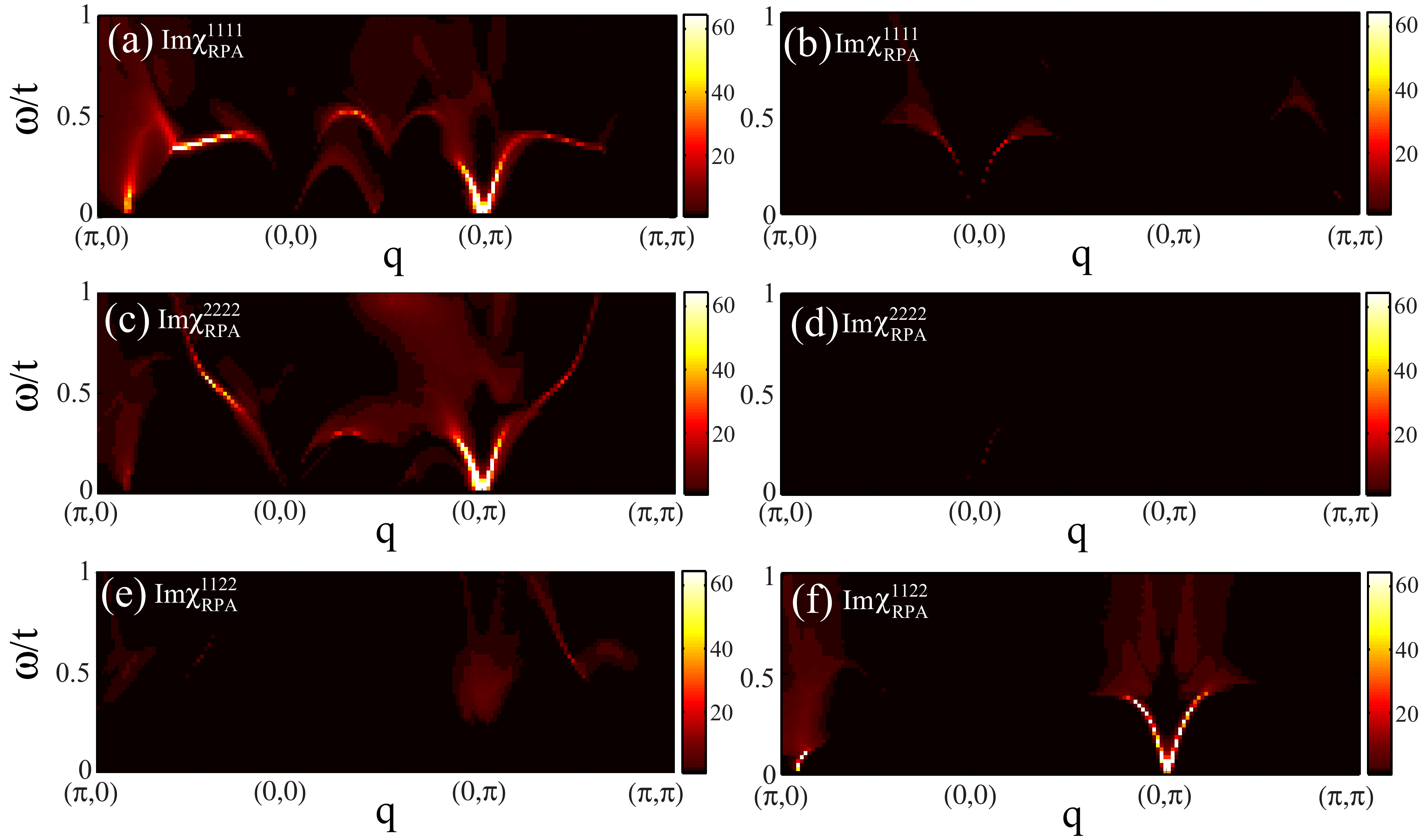}
\caption{Components of the transverse RPA spin susceptibility for the orbital (left panel) and excitonic (right panel) scenarios for a magnetic moment of $m=0.38 \mu_B$. The intensity is shown on an absolute scale.}
\label{fig5}
\end{figure}

A peculiarity of the spin waves within the orbital scenario is the relatively rich structure of the spin excitations
away from the ordering momentum. Here the complication is that each orbital has its own magnetic gap which correspondingly affects the magnetic excitations within each orbital. In Fig.\ref{fig5} we display the intra and inter-orbital components of the transverse spin susceptibility for the orbital and the excitonic scenario of the SDW phase on an absolute intensity scale.
Observe that within the orbital scenario, the Goldstone modes originate from the intra-orbital $xz$ and $yz$ susceptibilities, while in the excitonic scenario the entire low energy excitations are in the inter-band susceptibility by construction. In addition, as one sees from the left panel, $(0,\pi)$ order breaks the symmetry between $xz$ and $yz$ orbitals which leads to different behavior of the spin excitations away from the ordering momentum and higher energies visible in the corresponding intra-orbital susceptibilities. Given the fact that the spin waves show some orbital dependences we believe that this feature could be measured by RIXS experiments\cite{Brink2010,Guarise2010}.

Another interesting feature of the spin excitations in the orbital scenario is that their spin wave branches reappear out of the continuum at higher energies, e.g. when going from $(0,0)$ to $(0,\pi)$ as shown in Fig.\ref{fig3}(c),(e). Here, the orbital coherence factors yielding the larger continuum (damping) are small at these higher energies, thus allowing the spin waves to reappear. As soon as the itinerant magnetism appears in MOS with multiple order parameters this reappearance of the spin waves at higher energies away from the ordering momentum  should be a general feature not restricted only to iron-based superconductors.

Spin excitations within the orbital scenario show a rich structure at energies which exceed well the onset of the particle-hole continuum. Note that at these energies the spin excitations should not any longer be interpreted as spin waves originating from breaking of the spin-rotational symmetry in the antiferromagnetic state but rather should be attributed to a Stoner continuum. These excitations are almost unchanged across the Neel temperature and as such can be well separated in the experimental data.

\begin{figure}[t]
\centering
\includegraphics[width=0.9\textwidth]{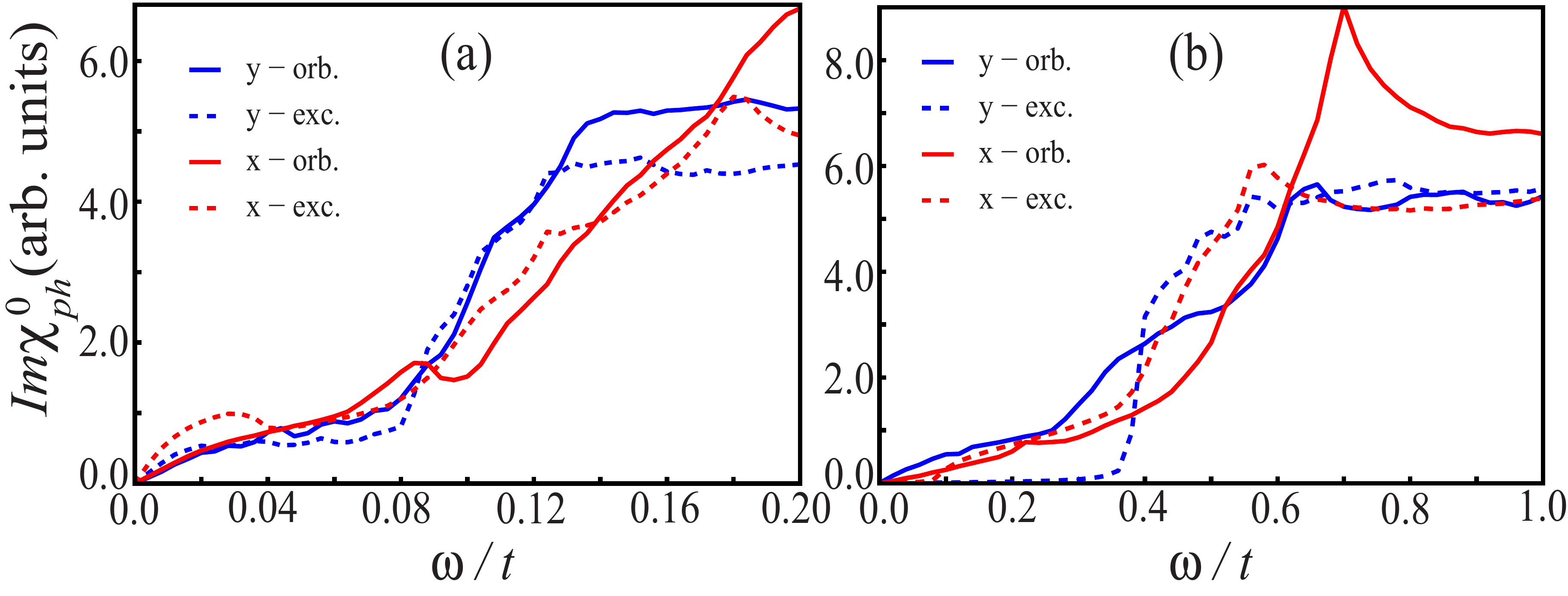}
\caption{Summed bare susceptibility versus energy for $m=0.23 \mu_B$ (a) and $m=0.38 \mu_B$ (b) in the antiferromagnetic y-direction (between $(0,0)$ and $(0,\pi)$) and the ferromagnetic x-direction (between $(0,\pi)$ and $(\pi,\pi)$) for the orbital (solid curves) and excitonic scenario (dashed curves).}
\label{fig6}
\end{figure}

In Fig.\ref{fig6} we show the behavior of the Im$\chi^{0}$ summed over the $q_x$ and $q_y$ momenta (compare to Fig.\ref{fig4}) for two different magnetic moments. Observe that for $m=0.22\mu_B$ (left) the damping increases slowly at low frequency and remains largely featureless up to energies of $\omega \sim 0.08t$. At higher frequencies damping becomes significantly larger and all excitations above this frequency refer to the Stoner continuum. For $m=0.38\mu_B$ (right) the continuum is gapped in the excitonic scenario at low frequencies but apart from this, the situation remains the same. In particular, at low enough frequencies, damping is weak and spin excitations can be associated with spin waves. For energies higher than $\omega \sim 0.4t$ the damping is so large that the excitations exist in  the form of the Stoner continuum only. The latter will only be weakly sensitive to the antiferromagnetic transition. Note that several experiments have studied the temperature dependence\cite{Harriger2010,Diallo2010} of the spin excitations and found that the high energy part of the spectrum is not sensitive to the magnetic or structural transition. In contrast to their interpretation of this result as local moments present at higher energies, we believe that the most natural explanation of these experiments should be given in terms of the Stoner continuum which despite its structure does not refer to spin wave excitations originating from the AF state but is mostly a property of the paramagnetic state. Therefore, it is almost unchanged across the transition.

Despite the widely known fact that the two orbital model is too simplistic to lead to quantitative results there are several aspects of our calculations which we believe will also exist in a more accurate  five orbital model. This includes the anisotropies of the spin wave velocities\cite{Diallo2009,Zhao2009}, occurrence of strong damping of the spin excitations in the orbital scenario with multiple order parameters and the hidden symmetries of the spin waves within the excitonic scenario which manifest themselves in accidental zero modes at certain non-ordering momenta like $(\pi,\pi)$ or $(\pi,0)$. We also find that for the nodal SDW, it is unavoidable to have a particle-hole continuum even at low energies away from the ordering wave vector independent of the particular details of the model. The puzzle wether or not damping of spin waves via the particle-hole continuum is observed can be resolved by the observation that in our calculations damping is always there even at low energies as long as some remnant portions of the FS are present. As seen in Fig.\ref{fig6}, the continuum itself can have additional steps and structure as a function of frequency.

Regarding whether the pure excitonic or orbital scenario has to be employed depends on the energies. For example, recently Ewings et al.\cite{Ewings2010} measured spin waves in $SrFe_2As_2$ throughout the whole BZ and up to high eneries. They conculded that a purely local moment model is not able to fit the data. By comparing excitonic scenario\cite{knolle} with the orbital one\cite{Tohyama_spinwaves} they found that  the low energy part of the spin wave dispersion can be well described by both scenarios, whereas, the orbital scenario gives generally a better description of the spin excitations at high energies. Most likely an inclusion of the momentum dependence of the interactions in the excitonic scenario would also improve the comparison.


\section{Summary}

Motivated by the magnetism present in the metallic iron-based superconductors we have analyzed spin excitations in magnetically ordered multiorbital systems within a self-consistent mean-field RPA-type formalism. We presented an extensive discussion of this theory and as an example carried calculations for the two orbital model. This formalism can be easily generalized to different MOS. As the magnetic electron-hole pairing can either reside on the orbital or at the band level, we have compared the behavior of spin excitations in two scenarios: either the on-site orbital one, or the excitonic scenario, where the magnetic order occurs on the band level in momentum space.
For small magnetic moments the low energy spin excitations look similar in both cases. In particular, the excitations are spin waves around the ordering momentum which become overdamped at higher energies away from {\bf Q}$_{AF}$. Due to remnant FS pockets the behavior of the particle-hole continuum is quite similar in both cases. The situation changes for larger magnetic moments, as the particle-hole continuum gets completely gapped within the excitonic scenario, while it remains by symmetry reasons gapless due to Dirac cones in the orbital description.
As a result in the excitonic scenario spin waves develop over the entire BZ with accidental zero modes at $(\pi,0)$ and $(\pi,\pi)$ points of the BZ due to an underlying accidental $O(6)$ symmetry of the spin excitations in this case. At the same time, in the orbital picture the spin waves remain well-defined only around the ordering momentum and are overdamped away from it. In addition we also find that due to multiple gaps in the orbital scenario the spin wave branches may reappear deeply inside the continuum because of the orbital character of the wave functions.
Analyzing the components of the spin susceptibility and the behavior of the particle-hole continuum within both scenarios one could clearly distinguish the orbital and the excitonic origin of the spin excitations in iron-based superconductors.

We thank A.V. Chubukov, P.M.R. Brydon, C. Timm, P. Dai, and T. Perring for useful discussions. The research of JK was supported by a Ph.D. scholarship from the
Studienstiftung des deutschen Volkes.
IE acknowledges the DAAD PPP Grant N50750339 for the financial support and MPI PKS for hospitality.


\begin{thebibliography}{99}
\bibitem{footnote1} Note that our previous study \cite{knolle} found that the other possible density-density interactions do not  change the structure of the spin waves qualitatively.
\end{thebibliography}
\end{document}